\documentclass[aip,jcp,reprint,amsmath,amssymb]{revtex4-1}
\usepackage{graphicx}
\usepackage{braket}
\usepackage{amsmath}
\usepackage{wasysym}
\usepackage{xcolor}
\usepackage{enumitem}
\begin{document}

% \preprint{}

\title{Long-range model of vibrational autoionization in core-nonpenetrating Rydberg states of NO} %Title of paper



\author{Timothy J. Barnum}
\email{tbarnum@mit.edu}
\affiliation{Department of Chemistry, Massachusetts Institute of Technology, Cambridge, Massachusetts 02139, USA}
%\thanks{}
\author{Gloria Clausen}
\affiliation{Laboratorium f\"ur Physikalische Chemie, ETH Z\"urich, Vladimir-Prelog-Weg 2, 8093 Z\"urich, Switzerland}
\author{Jun Jiang}
\altaffiliation{Present address: Lawrence Livermore National Laboratory, 7000 East Avenue, Livermore, California 94550, USA}
\author{Stephen L. Coy}
\author{Robert W. Field}
\altaffiliation{Author to whom correspondence should be addressed: rwfield@mit.edu}
\affiliation{Department of Chemistry, Massachusetts Institute of Technology, Cambridge, Massachusetts 02139, USA}


\date{\today}

\begin{abstract}

In high orbital angular momentum ($\ell \geq 3$) Rydberg states, the centrifugal barrier hinders close approach of the Rydberg electron to the ion-core. As a result, these \textit{core-nonpenetrating} Rydberg states can be well described by a simplified model in which the Rydberg electron is only weakly perturbed by the long-range electric properties (i.e., multipole moments and polarizabilities) of the ion-core. We have used a long-range model to describe the vibrational autoionization dynamics of high-$\ell$ Rydberg states of nitric oxide (NO). In particular, our model explains the extensive angular momentum exchange between the ion-core and Rydberg electron that had been previously observed in vibrational autoionization of $f$ ($\ell=3$) Rydberg states. These results shed light on a long-standing mechanistic question around these previous observations, and support a direct, vibrational mechanism of autoionization over an indirect, predissociation-mediated mechanism. In addition, our model correctly predicts newly measured total decay rates of $g$ ($\ell=4$) Rydberg states because, for $\ell\geq4$, the non-radiative decay is dominated by autoionization rather than predissociation. We examine the predicted NO$^+$ ion rotational state distributions generated by vibrational autoionization of $g$ states and discuss applications of our model to achieve quantum state selection in the production of molecular ions.

\end{abstract}


\maketitle %\maketitle must follow title, authors, abstract and \pacs


\section{Introduction}

Among the most compelling new topics at the intersection of chemistry and physics is the study of chemical reactions and molecular collisions at extremely cold temperatures in which quantum state specific behavior is resolved.\cite{stuhl2014, heazlewood2021, toscano2020} A diverse range of experimental techniques encompassing beams, traps, and cryogenic matrices now enable study of collision energies from a few K to sub-mK (see Ref.\ \citenum{toscano2020} and references therein). Beyond control of the collision energy, resolution of individual quantum states of both reactants and products has led to some of the most detailed pictures of chemical reactions to date.\cite{liu2020, hu2021} In such quantum-state-resolved studies of chemical reactivity, the generation of reactant molecules or ions in single quantum states is one of the greatest challenges and only a few systems have been experimentally realized thus far.\cite{tong2012, hu2019}

One approach to the preparation of quantum-state-selected samples is to selectively excite a Rydberg state of a neutral molecule, which then autoionizes to produce a molecular ion in the desired quantum state. Autoionization refers to the spontaneous ionization of a neutral Rydberg molecule when the total energy of a Rydberg electron around a vibrationally excited ion-core exceeds the ionization energy of the vibrationless level in one or more rotational states. Naively, this process would be expected to leave the ion-core rotational state unchanged because of the mismatch of vibronic and rotational time scales. However, in recent applications, this approach has only succeeded in producing molecular ions distributed over at least three final rotational states.\cite{zhou2019,loh2011} Indeed, earlier work demonstrated that rotational quantum number changes as large as four quanta are possible in the autoionization of certain Rydberg series,\cite{park1996,park1997} upending the simple picture of the Rydberg electron being released without perturbation of the ion-core.

In a series of pioneering experiments on the Rydberg states of nitric oxide (NO), the Zare group\cite{park1996,park1997,konen2003, zhao2004_thesis} employed photoelectron spectroscopy to examine the NO$^+$ ion rotational state distributions following vibrational autoionization of a selected Rydberg state. In nearly all cases, the autoionization of $s$, $p$, and $f$ Rydberg states of NO resulted in extensive angular momentum exchange between the Rydberg electron and the ion-core. Moreover, the observation of both even and odd parity final rotational states imply ejection of the Rydberg electron with both even and odd values of $\ell$, a surprising result in light of the small dipole moment of NO which induces only weak interactions between Rydberg series of even and odd $\ell$.\cite{miescher1976,fredin1987,cheung1983} It was concluded that at least two mechanisms are required to explain this new data: (1) direct coupling of the bound Rydberg state to even- and odd-$\ell$ channels in the ionization continuum by vibrationally dependent matrix elements and (2) indirect, electronic coupling to the ionization continuum through dissociative valence states.\cite{park1997} A later study\cite{pratt1998} suggested that the indirect coupling mechanism alone can explain the production of both even- and odd-$\ell$ photoelectrons as a result of the strong Rydberg-valence coupling responsible for the rapid predissociation of most NO Rydberg states. To date, no quantitative theory has accounted for the mechanisms operative in vibrational autoionization of NO.

In the present work, we apply a long-range model of vibrational autoionization to account for the dynamics of $f$ ($\ell=3$) and $g$ ($\ell=4$) Rydberg states of NO. The model relies on the simplifying assumption that the interaction of the Rydberg electron with the ion-core occurs only at long range via the electric properties of the ion-core.\cite{russek1968, eyler1986} This assumption explicitly limits the applicability of the long-range model to \emph{core-nonpenetrating} Rydberg states: states with sufficiently high $\ell$ (typically, $\ell \geq 3$) such that the Rydberg electron has minimal overlap with the spatial region of the ion-core. In spite of its simplicity, this model accounts for the majority of observed rotational decay channels following autoionization of $f$ states, providing explicit support for the proposed direct, vibrational autoionization mechanism.

We have measured total decay rates for autoionizing $g$ levels and find good agreement between those measurements and our model predictions of autoionization rates in the absence of predissociation. The success of our long-range model in explaining these experimental results has led us to a theoretical examination of the rotational state distribution of autoionizing $g$ Rydberg states. We find production of a single ($>$90\%) rotational state of the NO$^+$ ion is possible via autoionization of selected $g$ Rydberg states. We discuss how this universal predissociation-free behavior of high-$\ell$ Rydberg states may be used for optimally state-selective production of molecular ions for cutting-edge applications in molecular physics and cold chemistry. 

\section{Theoretical and Computational Methods}
\label{theorysection}
The physical basis of long-range autoionization in molecular Rydberg states has been described previously,\cite{russek1968, eyler1986} and has been demonstrated to accurately predict the autoionization rates of $f$, $v=1$ Rydberg states of H$_2$.\cite{eyler1986,lindsay1990} The autoionization rate can be obtained from Fermi's golden rule:
\begin{equation}
\Gamma = \dfrac{2\pi}{n^3}\left|\left<\Phi_f\left|\mathbf{H^\prime
}\right|\Psi_i\right>\right|^2 \label{fermisgoldenrule}
\end{equation}
where \textbf{H$^\prime$} is the Hamiltonian that describes all long-range interactions, $\Phi_f$ is the final continuum state, and $\Psi_i$ is the initial Rydberg state. In the results presented here, we explicitly consider contributions from the dipole moment, quadrupole moment, and dipole-dipole polarizability of the NO$^+$ ion-core:
\begin{equation}
    \mathbf{H^\prime
}=\mathbf{H_{dipole}}+\mathbf{H_{quad}}+\mathbf{H_{pol}}
\end{equation}
The initial Rydberg state is specified by quantum numbers for the vibrational state of the ion-core ($v$), the rotational state of the ion-core ($R$), the principal quantum number ($n$), the Rydberg electron orbital angular momentum ($\ell$), and the total angular momentum excluding spin ($N$) and its space-fixed projection ($M$). The final continuum state is specified by the vibrational ($v^+$) and rotational ($N^+$) quantum numbers for the bare ion, the energy above threshold ($\epsilon$) and orbital angular momentum ($\ell^\prime$) of the electron, and the total angular momentum excluding spin of the system ($N^\prime$) and its space-fixed projection ($M^\prime$). Selection rules for all such interaction mechanisms require $N^\prime=N$ and $M^\prime=M$. All matrix elements can be computed following the method of Eyler and Pipkin.\cite{eyler1983} For example, the multipolar interaction of rank $k$ has the form:
\begin{align}
&\left<v^+N^+\epsilon\ell'N'M'|\mathbf{H_{k}}|vRn\ell NM\right> \nonumber \\
& \quad =  -e\left<v^+N^+|Q_k(z)|vR\right>\left<\epsilon\ell'|r^{-(k+1)}|n\ell\right> \\
& \qquad \times \delta_{N',N}\delta_{M',M}(-1)^{\ell+\ell'+N} \nonumber \\
& \qquad \times \left[(2\ell'+1)(2\ell+1)(2N^++1)(2R+1)\right]^{1/2} \nonumber \\
& \qquad \times \begin{Bmatrix}N&N^+&\ell'\\k&\ell&R\end{Bmatrix}\begin{pmatrix}\ell'&k&\ell\\0&0&0\end{pmatrix}\begin{pmatrix}N^+&k&R\\0&0&0\end{pmatrix} \label{dipolematrixelement}
\end{align}

The selection rules for changes in the rotational ($R$ and $N^+$) and orbital ($\ell$ and $\ell^\prime$) angular momentum quantum numbers are determined by the Wigner 3-j symbols. For the mechanisms considered in our model:
\begin{itemize}[leftmargin=*]
    \item[] dipole: $\Delta\ell=\pm1; N^+-R=\pm1$
    \item[] quadrupole: $\Delta\ell=0,\pm2; N^+-R=0,\pm2$
    \item[] isotropic polarizability: $\Delta\ell=0; N^+-R=0$
    \item[] anisotropic polarizability: $\Delta\ell=0,\pm2; N^+-R=0,\pm2$
\end{itemize}

In Eq.\ \ref{dipolematrixelement}, $Q_k(z)$ represents the $k^{\mathrm{th}}$ multipole moment of the ion-core as a function of the internuclear distance $z$. In the harmonic approximation, interactions via this term are dominated by changes of $\Delta v=-1$. We have calculated the potential energy curve (PEC) and the electric properties of the NO$^+$ ion as a function of internuclear distance using the \texttt{ORCA} program suite.\cite{neese2012, neese2017} Geometry optimization calculations at 0.1 a.u.\ intervals in the region of the equilibrium internuclear distance were performed employing the Complete Active Space Self Consistent Field (CASSCF) method, followed by N-valence electron perturbation theory (NEVPT2) with the aug-cc-pVQZ basis set. The use of NEVPT2 to capture dynamical correlation effects significantly improved the PEC of NO$^+$ relative to a previously reported calculation.\cite{feher1993} Vibrational wave functions were computed on this PEC by a 1-D Discrete Variable Representation (DVR) scheme\cite{colbert1992} and used to calculate the matrix elements for electric properties.

The second term in Eq.\ \ref{dipolematrixelement} involves radial electronic wavefunctions for both the bound $\ket{n\ell}$ and continuum $\ket{\epsilon\ell^\prime}$ states of the Rydberg electron. These were assumed to be hydrogenic and were generated numerically by Numerov integration on the Coulomb potential using square-root scaling of the radial distance $r$ from the ion-core.\cite{bhatti1981} The continuum wavefunctions were normalized to the value of the analytic wavefunction at the position of its first maximum computed using the generalized hypergeometric function implemented in the \texttt{MATLAB} software package. Both the continuum and bound Rydberg state wavefunctions are energy normalized. For the bound states, this introduces a factor of $n^{1.5}$ in the calculation of $\mathbf{H^\prime}$, which when squared cancels with the factor of $n^{-3}$ in Eq.\ \ref{fermisgoldenrule}. When using this energy normalization, the squared radial matrix elements, representing the various long-range mechanisms, approach nearly constant values as $n$ increases. Thus, the calculated autoionization rates display nearly exact $n^{-3}$ scaling, in agreement with the experimental data presented later in Figure \ref{fig:ratedata_summaryfigure}. For this reason, we have designated the $n^{1.5}$ scaled interaction Hamiltonian as $\mathbf{H^\prime}$ and made the universal $n^{-3}$ scaling of the autoionization rates explicit in Eq.\ \ref{fermisgoldenrule}.

In calculating total autoionization rates, interference effects between the various long-range mechanisms are considered by summing the amplitudes of the different mechanisms before computing the autoionization rate according to Eq.\ \ref{fermisgoldenrule}. Since different final rotational states of the ion represent physically distinct decay channels, we can compute the autoionization yield of a particular ion rotational state by summing over the autoionization rates for all values of the final electron angular momentum $\ell^\prime$. Our treatment neglects interference among outgoing electron partial waves with different values of $\ell^\prime$ and thus the photoelectron angular distributions are not accessible by our calculations. This limitation results in a quantitatively inexact comparison between our theoretical results and some of the experimental photoelectron spectroscopy results,\cite{park1996,park1997} as discussed below.

Throughout this work, we specify individual Rydberg states using the compact notation $n\ell R_N$, where the quantum numbers are defined as above. In some cases, it is convenient to specify a particular electric fine structure component by the label $\ell_R=N-R$ rather than $N$. This label facilitates comparison of Rydberg states with different values of the core rotation, but the same relative coupling between rotational and electronic angular momentum. 

\section{Experimental Methods}

As an additional point of validation for our theoretical results, we sought to measure the total decay rates of $g$ series Rydberg states. Rydberg states with $\ell=4$ must be accessed via an intermediate state with $\ell=3$ due to a $\Delta\ell=\pm1$ selection rule for transitions between high-$\ell$ molecular Rydberg states. For NO, we employed a previously described triple resonance excitation scheme that uses the A $^2\Sigma^+$ and $4f$ levels as intermediate states.\cite{fujii1995} The laser radiation for the triple resonance excitation scheme was generated by three pulsed dye laser systems. The third harmonic (355 nm) of a pulsed Nd:YAG laser (Spectra Physics GCR 270) pumped two pulsed dye lasers (SIRAH Cobra Stretch, Lambda Physik Scanmate 2E), both operated with Stilbene 420 dye. The output of the SIRAH dye laser was frequency doubled in a $\beta$-barium borate crystal to generate the radiation around 217 nm appropriate for pumping rotational transitions in the A $^2\Sigma^+$ ($v=1$) $\leftarrow$ X $^2\Pi_{1/2}$ ($v=0$) band. The fundamental output of the Lambda Physik dye laser in the 420 nm region was resonant with transitions in the $4f$ ($v=1$) $\leftarrow$ A $^2\Sigma^+$ ($v=1$) band. The intensity for this transition is derived from the partial $d$ ($\ell=2$) character of the A $^2\Sigma^+$ state, as discussed previously.\cite{cheung1983} The IR radiation around 1500 nm needed to drive $ng$ ($v=1$) $\leftarrow 4f$ ($v=1$) transitions was generated by a third pulsed dye laser (Continuum ND6000), operated with DCM dye and pumped by the second harmonic (532 nm) of a pulsed Nd:YAG laser (Spectra Physics Pro 270). The output of this dye laser was then mixed with the fundamental (1064 nm) of the same Nd:YAG laser in a LiNbO$_3$ crystal to produce IR radiation by difference frequency generation. All three laser beams were overlapped with dichroic mirrors and loosely focused into the detection chamber.

We measured the total decay rates of $g$ ($v=1$) Rydberg states, over a range of values of $n$, $R$, and $\ell_R$, by delayed pulsed field extraction of NO$^+$ ions resulting from autoionization. The ion detection apparatus has been described in detail previously\cite{kay2004} and only relevant details are given here. NO gas seeded at 0.5 \% in Ar was expanded  through a pulsed valve (General Valve Series 9, $d=1$ mm) and passed through a conical skimmer ($d=0.5$ mm) before entering the differentially pumped detection chamber that contains a Wiley-McLaren type time-of-flight mass spectrometer. The molecular beam was crossed at right angles by the three excitation lasers. By strongly attenuating the first and second lasers and adding 10 ns delays between each of the laser pulses, the ion signal due to multiphoton ionization was minimized. Following a variable delay time after the third laser pulse, a pulsed field of 200 V/cm was applied to the bottom electrode and ions produced by vibrational autoionization were accelerated through a 75 cm field-free region and impinged on a  multichannel plate detector. The pulsed field was weaker than that required to field-ionize the Rydberg states investigated in this work. An exponential function was fit to the raw ion yield data starting approximately 5 ns after the first appearance of ions to minimize the influence of the extraction field rise time. The analysis of this data is discussed in more detail in Section \ref{grates_discussion}.

Two experimental effects can interfere with the described measurements. First, stray electric fields can mix $g$ states with $f$ states and high-$\ell$ Rydberg states, altering the lifetimes relative to those in field-free conditions.\cite{vrakking1995,murgu2001} We estimate the stray electric field in our apparatus by measuring the $44h \leftarrow 43g$ transition at approximately 80 GHz by broadcasting a weak microwave field into the detection region following laser population of the initial $43g$ state. Based on the observed line broadening, we estimate the presence of a stray electric field of approximately 90 mV/cm in the apparatus during these experiments. We performed a Stark effect calculation analogous to that presented by \citeauthor{vrakking1996}.\cite{vrakking1996} With this stray electric field, we found that, for the highest $n$ values investigated ($n=28$), the purity of all states was greater than 97\% and the lifetimes were altered by less than 3\% relative to those of the field-free states.

Second, the laser linewidth in the final excitation step of our experiment was approximately 0.05 cm$^{-1}$, substantially larger than the electric fine structure splitting of the $g$ Rydberg states. For some levels ($ng0_4$, $ng1_3$, $ng2_2$, and $ng3_1$, where the quantum numbers $n\ell R_N$ are defined in Section \ref{theorysection}), we could unambiguously excite each of these selected electric fine structure components by appropriate selection of the intermediate levels. For all other levels where perfect fine structure selection was not possible, we took advantage of a strong $\Delta N=\Delta\ell=+1$ propensity rule in the final step of the excitation scheme. Thus, by careful selection of the intermediate state we could \textit{predominantly} excite a single $\ell_R=N-R$ component of the Rydberg state despite our inability to resolve individual states. We determined the anticipated purity of the final state preparation by explicitly calculating the triple-resonance transition intensities. Following the procedure of \citeauthor{petrovic2008},\cite{petrovic2008} we  first calculated the transition intensity for each step in a Hund's case (a) basis and then performed a rotational frame transformation to the correct Hund's case for each level. This treatment, which explicitly considers the summation over unresolved spin doublets, has previously been shown to predict accurate relative intensities in multiple resonance experiments.\cite{petrovic2008} 

In Table \ref{table:excitationschemes}, we summarize the excitation schemes and the calculated selectivities of each excitation scheme for every level investigated in this work. For the majority of states, very high purity of the desired state is predicted. However, the selectivities for $ng3_2$, $ng3_3$, and $ng3_5$ states are only 0.63, 0.61, and 0.72, respectively. The majority of the contamination in this preparation scheme is from the unresolved $Q$-type transition to the electric fine structure component with total angular momentum $N-1$ relative to the desired $N$ level. In spite of this contamination, it is evident from the results presented later in Figure \ref{fig:ratedata_summaryfigure} and discussed in Section \ref{grates_discussion}, that these less-perfectly selected states do not exhibit greater discrepancies from the calculation than the other data. This may simply be a result of the modest variation in rates expected among the different $\ell_R$ states, so that contamination of the desired state preparation by a neighboring state will not dramatically alter the measured rates.

\begin{table}
	\caption{Excitation scheme and selectivity for preparation of the indicated Rydberg state. In all cases, a final $R$-type transition to the desired $N$ level is the most intense line. Nearly all contamination is due to a $Q$-type transition to the $N-1$ component. The A$\leftarrow$X transitions are labeled as $\Delta J_{F_{1/2} F_{1/2}\prime}$ where the number subscripts indicate a transition between the $F_1$ spin component of the X state, namely X $^2\Pi_{1/2}$ levels, and the $F_1$ or $F_2$ spin component of the A state, namely $J=N+1/2$ or $J=N-1/2$ levels with the same value of $N$. \cite{herzberg1950} The $4f\leftarrow$A transitions are labeled as $^{R-N'}\Delta N_{\ell_R}$ where $N'$ is the total angular momentum excluding spin of the A state level. \cite{cheung1983}}
	\centering
	\setlength{\tabcolsep}{8pt}
	\begin{tabular}{ c c c c}
	\hline\hline 
	State & A$\leftarrow$X &  $4f\leftarrow$A &  Selectivity \\ [1ex]
	\hline 						
	$ng0_4$ & R$_{21}$(0.5) & $^{-2}$R$_3$(2) & 1.0  \\ 
	$ng1_5$ & R$_{21}$(1.5) & $^{-2}$R$_3$(3) & 0.988 \\
	$ng1_4$ & R$_{21}$(1.5) & $^{-2}$Q$_2$(3) & 0.976 \\
	$ng1_3$ & P$_1$(2.5) & $^0$R$_1$(1) & 1.0 \\ 
    $ng2_6$ & R$_{21}$(2.5) & $^{-2}$R$_3$(4) & 0.972 \\
	$ng2_5$ & R$_{21}$(2.5) & $^{-2}$Q$_2$(4) & 0.920 \\
	$ng2_4$ & R$_{21}$(0.5) & $^0$R$_1$(2) & 0.965 \\
	$ng2_3$ & R$_{21}$(0.5) & $^0$Q$_0$(2) & 0.959 \\	
	$ng2_2$ & P$_1$(1.5) & $^2$R$_{-1}$(0) & 1.0 \\
	$ng3_7$ & R$_{21}$(3.5) & $^{-2}$R$_{3}$(5) & 0.955\\
	$ng3_6$ & R$_{21}$(3.5) & $^{-2}$Q$_{2}$(5) &  0.872\\
    $ng3_5$ & R$_{21}$(3.5) & $^{-2}$P$_{1}$(5) & 0.724\\
	$ng3_3$ & R$_{21}$(1.5) & $^{0}$P$_{-1}$(3) & 0.614\\
	$ng3_2$ & P$_1$(2.5) & $^2$Q$_{-3}$(1) & 0.626 \\	
	$ng3_1$ & P$_1$(2.5) & $^2$P$_{-3}$(1) & 1.0 \\ [0.5ex]
	\hline
\end{tabular}
\label{table:excitationschemes}
\end{table}

\section{Results and Discussion}

\subsection{Rotational state distributions from autoionization of $f$ Rydberg states}

We calculated the NO$^+$ ion rotational state distributions that result from vibrational autoionization of selected $f$ Rydberg states by summing the rates of all outgoing electron partial waves that result in formation of a given final rotational state of the ion. The model predictions are compared to the experimental results for $v=1$  and $v=2$ $f$ Rydberg states from Ref.\ \citenum{park1997} and Ref.\ \citenum{zhao2004_thesis}, respectively. Briefly, these experiments prepared a selected $f$ Rydberg state by double resonance excitation of NO in a molecular beam. Vibrational autoionization of these Rydberg states occurred under field-free conditions in a time-of-flight photoelectron spectrometer. The energy resolution of the spectrometer was sufficient to resolve individual ion rotational levels for rotational quantum numbers $N^+\gtrsim13$. The laser resolution was insufficient to selectively prepare individual $\ell_R$ components of a Rydberg complex, resulting in simultaneous excitation of as many as three $\ell_R$ components. In most cases, photoelectron spectra were observed at a single angle between the time-of-flight axis and the polarization direction of the laser beams. In a few cases for $v=2$ Rydberg states, the extreme $\ell_R$ components ($\ell_R=\pm3$) were selectively excited by using circularly polarized laser light.\cite{zhao2004_thesis} In those cases, complete photoelectron angular distributions were also collected by observing the photoelectron spectra at several angles between the time-of-flight axis and the laser polarization axis. 

For both the $v=1$ and $v=2$ data sets, we examined only the $\Delta v=-1$ autoionization channel. The much slower $\Delta v=-2$ decay is likely contaminated by additional mechanisms such as the predissociation-mediated autoionization considered previously.\cite{giustisuzor1984} To clarify the distinction between the quantum numbers of the ion-core and the bare ion, we refer to the rotational quantum number of the ion-core as $R$ and the bare ion rotational quantum number as N$^+$, which is identical to the total angular momentum for the $^1\Sigma^+$ ground electronic state of NO$^+$.

\begin{figure}
\centering
\includegraphics{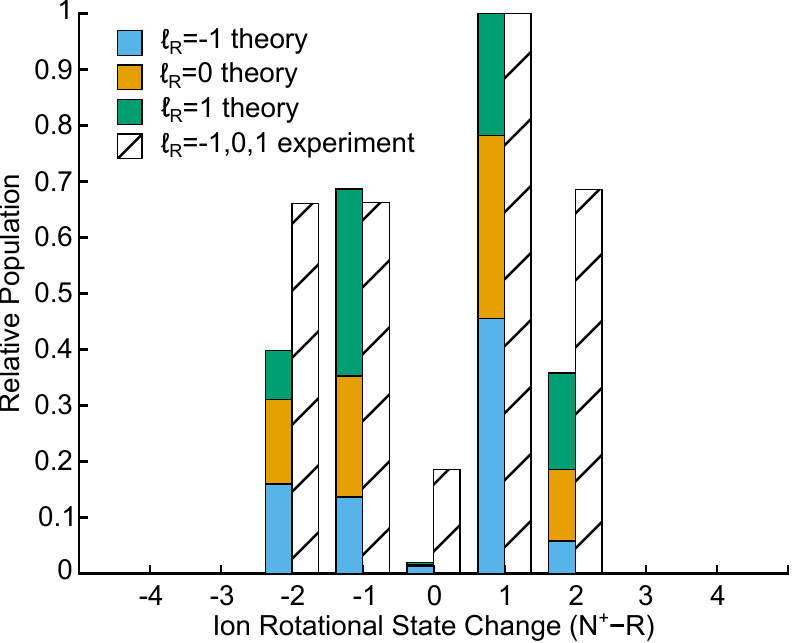}
\caption{Rotational state distribution following vibrational autoionization of the unresolved $13f16_{15,16,17}$ ($v=2$) states. The experimental data from Ref.\ \citenum{zhao2004_thesis} (hatched bars) are the photoelectron peak intensities observed at one angle, and serve as proxies for the true NO$^+$ ion rotational state distribution. The theoretical predictions appear as stacked bars in blue, orange, and green for the $\ell_R =$ -1, 0, and 1 states, respectively. Both data sets have been normalized by setting the most intense peak to 1. The long-range model predicts intensity in all five experimentally observed decay channels.}
\label{fig:v2ellR-101}
\end{figure}

Figure \ref{fig:v2ellR-101} shows a comparison between the experimental data and our model predictions for the $13f16_{15,16,17}$ ($v=2$) states. These three electric fine structure states have $\ell_R$ values of $-1$, 0, and 1. The relative intensities of the observed photoelectron peaks at energies corresponding to the given change of rotational state ($N^+-R$) are shown as hatched bars. The model predictions are shown as blue, orange, and green stacked bars for the $\ell_R=-1$, 0, and 1 states, respectively. It is important to note that the intensity of the photoelectron peak observed at a single angle does not precisely represent the yield for that rotational channel. The total yield is given by the photoelectron intensity summed over the full photoelectron angular distribution. The long-range model could be extended to predict photoelectron angular distributions by explicitly considering the interferences between several outgoing partial waves and by summing over the amplitudes of these exit channels. In this work, the photoelectron peaks observed at a single angle serve as proxies for the true NO$^+$ ion rotational state distribution.

Figure \ref{fig:v2ellR-101} demonstrates that the long-range model correctly predicts intensity in all five rotational state channels observed experimentally and reproduces the qualitative intensity pattern. Although a quantitative comparison of the theoretical and experimental intensities is not possible due to the limitations discussed above, it is important to note the significantly reduced intensity of the $N^+-R=0$ channel in both the experimental and theoretical results. Our mechanistically explicit model offers an explanation for this surprisingly weak decay channel. We find that vibrational autoionization into the $\Delta\ell=0$, $N^+-R=0$ channel occurs via the quadrupole and polarizability mechanisms with similar magnitudes, but opposite signs. Thus, an interference effect between these mechanisms suppresses the decay rate into this rotational state channel.

\begin{figure}
\centering
\includegraphics{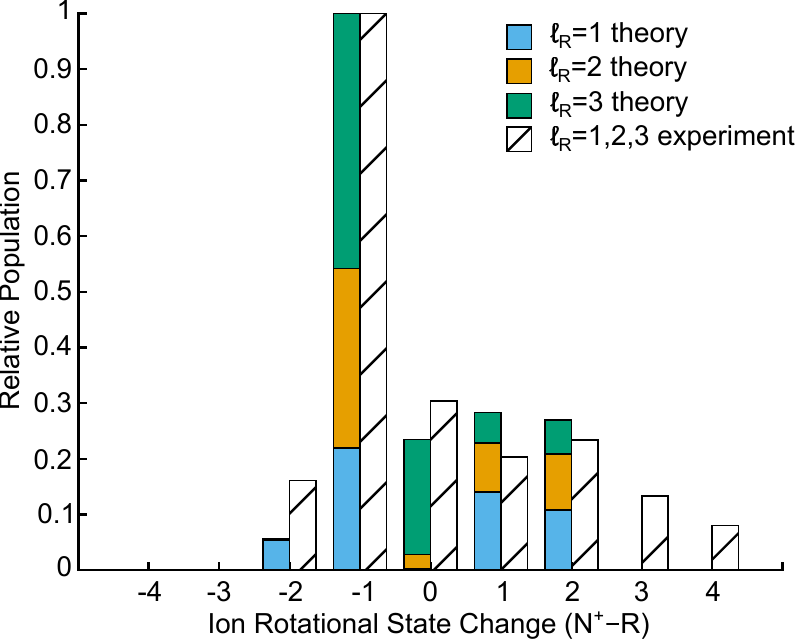}
\caption{Rotational state distribution following vibrational autoionization of the unresolved $13f17_{18,19,20}$ ($v=1$) states. The experimental data from Ref.\ \citenum{park1997} (hatched bars) are the photoelectron peak intensities observed at a single angle, and serve as proxies for the true NO$^+$ ion rotational state distribution. The theoretical predictions appear as stacked bars in blue, orange, and green for the $\ell_R =$ 1, 2, and 3 states, respectively. Both data sets have been normalized by setting the most intense peak to 1. The long-range model predicts a qualitatively correct intensity pattern for ion rotational state changes between $N^+-R=-2$ and $N^+-R=2$, but fails to account for the large rotational state changes of $N^+-R=3$ and 4.}
\label{fig:v1ellR123}
\end{figure}

\begin{figure}
\centering
\includegraphics{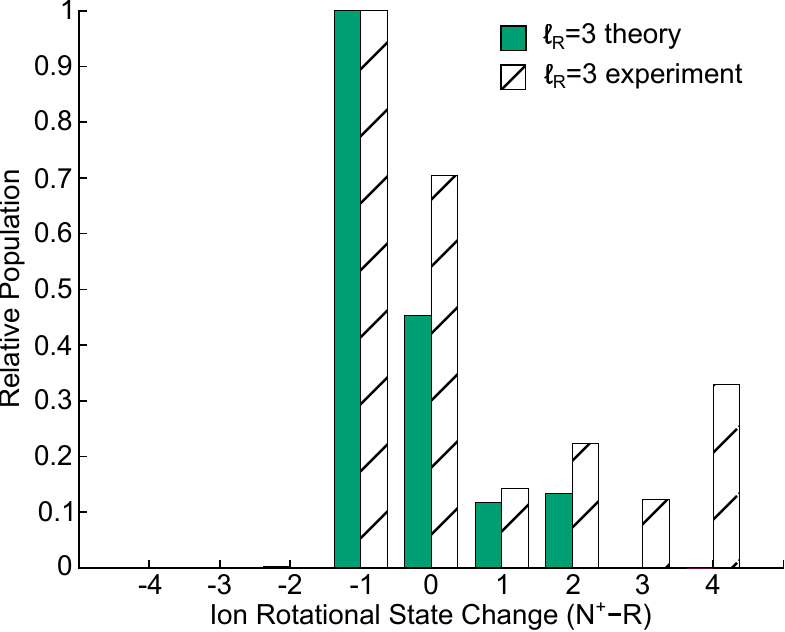}
\caption{Rotational state distribution following vibrational autoionization of the single $11f18_{21}$ ($v=2$) state. The experimental data from Ref.\ \citenum{zhao2004_thesis} (hatched bars) are the $\beta_{00}$ parameters from a fit to the measured photoelectron angular distributions. These values are directly proportional to the NO$^+$ ion rotational state distribution. The theoretical predictions appear in green. Both data sets have been normalized by setting the most intense peak to 1. The long-range model predicts the qualitatively correct intensity pattern for ion rotational state changes between $N^+-R=-2$ and $N^+-R=2$, but fails to account for the large rotational state changes of $N^+-R=3$ and 4.}
\label{fig:v2ellR3}
\end{figure}

Figure \ref{fig:v1ellR123} displays the rotational state distributions following vibrational autoionization for the $13f17_{18,19,20}$ ($v=1$) states, corresponding to $\ell_R=1$, 2, and 3, and Figure \ref{fig:v2ellR3} displays data for the single $11f18_{21}$ ($v=2$) state, corresponding to $\ell_R=3$. For the $\ell_R=3$ state in Figure \ref{fig:v2ellR3}, the plotted experimental data are the $\beta_{00}$ parameters from a fit to the measured photoelectron angular distributions, which are directly proportional to the NO$^+$ ion yield in that rotational channel. Rotational state distributions for the $\ell_R=-1,-2,$ and $-3$ states and the single $\ell_R=-3$ state have also been analyzed. In both the experimental and theoretical results, these states simply show a mirror-image rotational state distribution about $N^+-R=0$ relative to that of the positive $\ell_R$ states and will not be discussed in further detail.

Our model reproduces the experimental intensity pattern observed in the $N^+-R$ channels between $-2$ and 2. The experimental data for the unresolved $\ell_R=1,2$, and 3 components display a dominant $N^+-R=-1$ peak, a weak $N^+-R=-2$ peak and similar intermediate intensity peaks for $N^+-R=0,1$, and 2. Our model captures this qualitative pattern, including the weak $N^+-R=-2$ peak, which can only be generated via autoionization of the $\ell_R=1$ component. The selectively excited $\ell_R=3$ data in Figure \ref{fig:v2ellR3} again displays a dominant $N^+-R=-1$ peak. In contrast to Figure \ref{fig:v1ellR123}, no intensity is observed in the $N^+-R=-2$ channel because the $\ell_R=1$ state is not populated by the laser excitation. In addition, the $N^+-R=0$ channel is significantly more intense than the $N^+-R=1$ and 2 channels; this pattern is also captured by our model.

Unlike the experimental and calculated data for $\ell_R=-1$, 0, and 1, the experimental results in Figures \ref{fig:v1ellR123} and \ref{fig:v2ellR3} show intensity in decay channels $N^+-R=3$ and 4. These decay channels are absent in our model simply because the selection rules for the considered long-range mechanisms (dipole, quadrupole, dipole-dipole polarizability) do not allow for these large changes in angular momentum. We have explored extensions to our model that could produce such large rotational state changes, including higher-order multipoles (octupole, hexadecapole), polarizabilities (dipole-quadrupole, quadrupole-quadrupole, dipole-octupole), and hyperpolarizabilities (dipole-dipole-dipole, dipole-dipole-dipole-dipole). The selection rules for these various mechanisms do allow for large angular momentum changes, and we found that some intensity can be produced in the $N^+-R=3$ and 4 channels. Moreover, the intensity patterns for $\ell_R=3$ and $\ell_R=-3$ states are mirror images of each other, in agreement with the experimentally observed pattern. However, the contributions of the higher-order multipoles, the octupole and hexadecapole, were two orders of magnitude lower than those of the dipole, quadrupole, and polarizability. This small contribution occurs in spite of the fact that the electronic dependence of the octupole scales with distance in the same way as the dipole-dipole polarizability ($\propto r^{-4}$). The origin of their weak influence on autoionization is the smaller magnitude of these multipole moments, or more accurately for vibrational autoionization, the smaller dependence on internuclear distance of these multipole moments around the NO$^+$ equilibrium bond length. 

\begin{figure}
    \centering
    \includegraphics{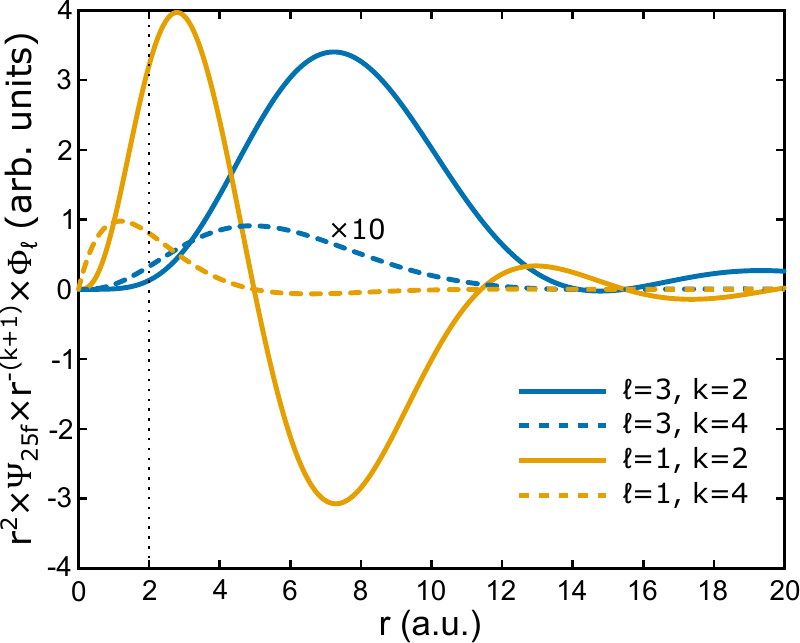}
    \caption{Integrands of the electronic matrix element as a function of radial distance for a $25f$ bound Rydberg state and continuum wavefunctions with $\epsilon=0.01$ and $\ell=1$ or 3. The integrand with $k=2$ corresponds to a quadrupole interaction, and $k=4$ to a hexadecapole interaction. The $\ell=3$, $k=4$ integrand has been scaled by a factor of 10 for visual clarity. The dotted vertical line at 2 a.u.\ represents the internuclear distance of NO$^+$ and approximates the boundary of the ion-core region. Low-$\ell$ outgoing waves contribute more at short radial distance where the long-range approximation becomes invalid. Higher-order electric properties (larger $k$) weight the integrand heavily at short radial distance, further driving the breakdown of the long-range approximation for low-$\ell$ outgoing waves.}
    \label{fig:integrand}
\end{figure}

In addition, we note that the contributions of the higher-order polarizabilities increased unexpectedly and unphysically as the mechanism becomes shorter in range (i.e., as the power $k$ of the radial matrix element $\braket{r^{-(k+1)}}$ increases). In some cases, the intensity pattern for $|N^+-R|<2$ channels was profoundly modified by the addition of a particular higher-order mechanism. We believe this unphysical behavior manifests a breakdown of the fundamental assumption of the long-range model: the Rydberg electron stays sufficiently far from the ion-core that short-range, many-electron interactions can be ignored. The unreasonable intensity of higher-order mechanisms predicted by our model is due mainly to large contributions from outgoing $\ell=0$ partial waves.  For outgoing electron waves with $\ell=0$, and to a lesser extent $\ell=1$, the long-range assumption is not valid. Figure \ref{fig:integrand} demonstrates this breakdown of the long-range approximation for low-$\ell$ outgoing waves. As the higher-order mechanisms become shorter in range, the integrand is weighted more heavily at small radial distance where low-$\ell$ waves contribute significantly. The NO$^+$ internuclear distance of approximately 2 a.u.\ is represented by a vertical dotted line in Figure \ref{fig:integrand} and approximates the boundary of the ion-core region, inside of which the long-range approximation is not valid. A phase shift of these low-$\ell$ waves due to close-range, many-electron physics will certainly occur and will significantly change contribution of these channels to autoionization. Since this physics is beyond the scope of our model, we neglect all higher-order electric properties. 

This limitation of the current model means that we cannot conclusively eliminate the role of an indirect, electronic mechanism in the vibrational autoionization of NO. Indeed, the $\ell_R=\pm3$ states have the greatest partial $\sigma$ character when projected into a Hund's case (b) basis set. Thus, these states might be expected to interact most strongly with the I $^2\Sigma^+$ and A$^\prime$ $^2\Sigma^+$ states that are predominantly responsible for predissociation\cite{pratt1998} and an indirect mechanism may contribute to decay channels with large rotational state changes. However, our model also clearly establishes that a direct, vibrational mechanism can account for the majority of the observed rotational decay channels, in spite of the rapid predissociation of $f$ Rydberg states. This is our most important observation. The faster rate of predissociation than autoionization for $f$ Rydberg states implies that the bound state interaction with the dissociation continuum  is stronger than with the ionization continuum.\cite{fujii1992,fujii1995} In an indirect autoionization path, a second interaction between the dissociation and ionization continua is required; our results imply that this continuum-continuum interaction is sufficiently weak as to make this indirect mechanism unimportant for most $f$ Rydberg states and decay channels studied here. Indeed, previous work points to the prominent role of indirect, predissociation-induced autoionization only in the case of $|\Delta v| > 1$ decay where the vibrational propensity rule predicts much slower direct, vibrational autoionization.\cite{giustisuzor1984} In future work, a more complete theory such as multichannel quantum defect theory\cite{jungen1997,jiang2019} will include both the short-range and long-range physics necessary to account for all of these different mechanisms.

\subsection{Total decay rates of $g$ Rydberg states}
\label{grates_discussion}
\begin{figure*}
\centering
\includegraphics{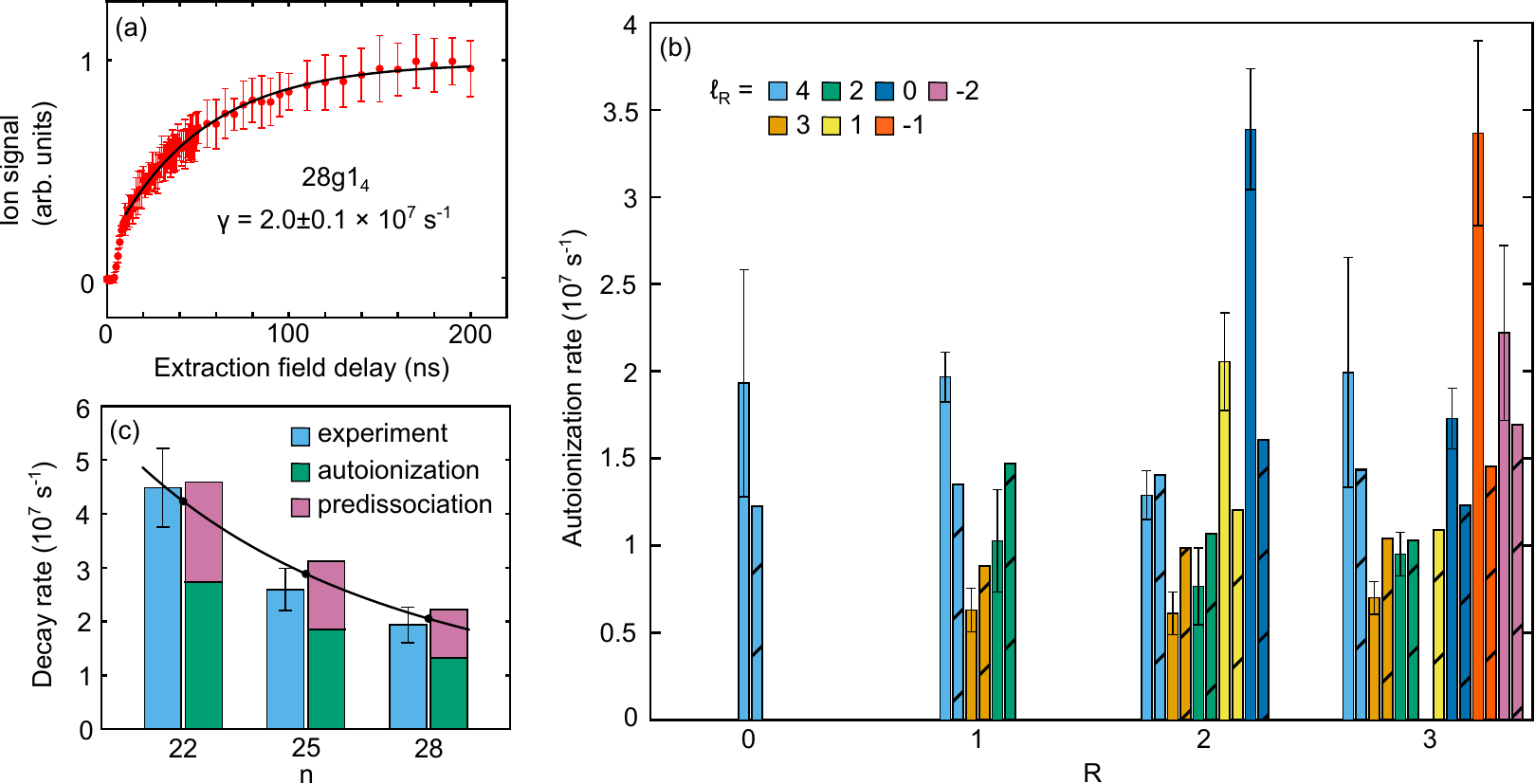}
\caption{(a) Ion signal as a function of extraction field delay time following excitation of the $28g1_4$ ($v=1$) state. Points are the average signal for 12 experiments and error bars represent the standard deviation of the 12 experiments. A fit to the data yields the experimental decay rate of this state, $\gamma$. (b) Experimental total decay rates (solid bars) and theoretical autoionization rates (hatched bars) for the $28g$ Rydberg states. States with different $R$ values are labeled on the abscissa and the possible $\ell_R$ values are shown as different colors with $\ell_R$ = 4 appearing on the left-hand side of each cluster. The error bars reflect the uncertainties of the fit to the raw experimental data. (c) The experimental decay rates for $22g$, $25g$, and $28g$, averaged over all $\ell_R$ and $R$ states investigated, are shown as light blue bars on the left-hand side with error bars representing an average uncertainty. The calculated autoionization rate, averaged again over all states, appears as the green bar to the right. A predissociation rate for each $n$ value, estimated from the work of Ref.\ \citenum{murgu2001}, appears as the violet bar stacked above the autoionization rate. The solid curve is a fit to the experimental data showing the expected $n^{-3}$ scaling of the decay rate.}
\label{fig:ratedata_summaryfigure}
\end{figure*}

Panel (a) of Figure \ref{fig:ratedata_summaryfigure} shows the measured ion signal as a function of the extraction field delay time following excitation of the $28g1_4$ Rydberg state. Delay times between 0 and 200 ns are randomized and sampled a total of 12 times for each data point. Error bars represent the standard deviation for the 12 measurements. The solid black line is a fit to the ion signal ($S$) with the functional form:
\begin{equation}
    S=S_{max}[1-e^{-\gamma(t-t_o)}]+S_o
\end{equation}
where $S_o$ and $t_o$ represent offsets in the initial signal and delay time, $S_{max}$ is the maximum observed ion signal at long delay times, and $\gamma$ is the total decay rate of the state. A fit offset of at least 5 ns from the onset of the ion signal was applied to avoid any influence from the turn-on of the extraction field. This influence is evident in the steeper slope of the data at very short delay times in Panel (a) of Figure \ref{fig:ratedata_summaryfigure} and is likely a result of exciting a Stark-mixed state with a faster decay rate. At later times when the extraction field has no overlap with the excitation laser, only $g$ state decay is probed because the ions present when the extraction field pulses on are focused on the MCP detector as a 10 ns wide, ``prompt" ion bunch. The remaining Rydberg population is Stark-mixed by the extraction field, which remains on for at least 1 $\mu$s in these experiments. Any ions produced from these states are spread out along the time-of-flight axis and not integrated into the ion signal that appears in Panel (a) of Figure \ref{fig:ratedata_summaryfigure}. All experimentally determined decay rates for the $g$ states measured in this work appear in Table \ref{table:lifetimes}.

\begin{table}
	\caption{Total decay rates of $g$ Rydberg states measured by delayed pulsed field extraction in this work, in units of $10^7$ s$^{-1}$. Numbers in parentheses are uncertainties ($2\sigma$) in units of the last reported digit.}
	\centering
	\setlength{\tabcolsep}{8pt}
	\begin{tabular}{ c c c c}
	\hline\hline 
	State & $n=22$ &  $n=25$ &  $n=28$ \\ [1ex]
	\hline 						
	$ng0_4$ & 7(2) & 0.4(2) & 1.9(7) \\ 
	$ng1_5$ & 3.1(5) & 1.5(2) & 2.0(1) \\ 
	$ng1_4$ & 1.9(3) & 1.0(1) & 0.6(1) \\
	$ng1_3$ & 4.6(7) & 2.6(3) & 1.0(3) \\
	$ng2_6$ & 2.2(5) & 3.0(4) & 1.3(1) \\
	$ng2_5$ & 1.0(1) & 0.9(2) & 0.6(1) \\
	$ng2_4$ & 2.4(5) & 0.9(3) & 0.7(2) \\
	$ng2_3$ & 4.7(9) & 2.0(3) & 2.1(3) \\
	$ng2_2$ & 5(1) & 3.4(6) & 3.4(4) \\	
	$ng3_7$ & 3.8(3) & 0.9(3) & 2.0(7) \\
	$ng3_6$ & 2.7(3) & 1.6(1) & 0.70(9) \\
	$ng3_5$ & 3.2(4) & 2.0(1) & 1.0(1) \\
    $ng3_3$ & 5.0(6) & 2.3(3) & 1.7(2) \\
	$ng3_2$ & 5.7(5) & 3.6(5) & 3.4(5) \\
	$ng3_1$ & 6(1) & 5.3(8) & 2.2(5) \\ [0.5ex]
	\hline
\end{tabular}
\label{table:lifetimes}
\end{table}

Panel (b) of Figure \ref{fig:ratedata_summaryfigure} shows the experimentally determined total decay rates (solid bars) and theoretical autoionization rates (hatched bars) for the $28g$ Rydberg state over a range of $R$ and $\ell_R$ values. The agreement between the experimental and theoretical values supports the conclusions of Ref.\ \citenum{fujii1995}: the non-radiative decay of $g$ ($v=1$) states is dominated by autoionization rather than predissociation. These results also stand in stark contrast with the behavior of the $f$ ($v=1$) states, which decay predominantly via predissociation at rates orders of magnitude faster than the predicted vibrational autoionization rates of our long-range model. Despite the imperfect selectivity of our preparation scheme, this data set also reveals a pattern of slower autoionization rates for the central $\ell_R$ components of a Rydberg complex (e.g., $\ell_R$=3 for the low rotational states in Figure \ref{fig:ratedata_summaryfigure}) which is in qualitative agreement with the model predictions. In the limit of high rotation, this pattern is explained by the very large decay rates of the $\Delta\ell=+1$, $N^+-R=\pm1$ channels for the extreme $\ell_R$ states. At the low-$R$ values shown in Figure \ref{fig:ratedata_summaryfigure}, the pattern is quantitatively different due to the presence of fewer states, but this qualitative pattern persists. Similar agreement between experiment and theory was found for $25g$ and $22g$ levels.

Panel (c) of Figure \ref{fig:ratedata_summaryfigure} summarizes the mechanisms contributing to the experimentally observed decay rates for the $n$=22, 25, and 28 states. The experimental data in light blue on the left-hand side is an average of the decay rates for all observed states with the given principal quantum number. To the right of the experimental data are stacked bars for the contributing decay mechanisms. The solid black curve is a fit to the experimental data, showing the $n^{-3}$ scaling of the average measured decay rate.

The radiative decay rate for these states is estimated from hydrogenic matrix elements;\cite{bethe1957} for $ng$ states, the decay scales as $5.54\times10^8$ $n^{-3}$ s$^{-1}$. At $n=28$, the radiative decay rate is estimated to be $2.5\times10^4$ s$^{-1}$, three orders of magnitude slower than autoionization and thus not visible on the scale of the plot. As pointed out in Ref.\ \citenum{eyler1986}, autoionization rates decrease more rapidly with $\ell$ than radiative decay rates and at high values of $\ell$, radiative decay will become the dominant mechanism. We find that for NO at $n=28$, this transition happens at $\ell=12$, where the estimated radiative decay rate and calculated autoionization rates are approximately equal at about $3\times10^3$ s$^{-1}$. 

The calculated autoionization rates, averaged over all observed states with that principal quantum number, appear as green bars to the right of the experimental data. The predissociation rate for $g$ states is adopted from Ref.\ \citenum{murgu2001} and consistent with the approximate lifetime of $\approx 1$ $\mu$s for the $n=55$ state measured in Ref.\ \citenum{fujii1995}. This predissociation rate scales as $1.97\times10^{11}$ $n^{-3}$ s$^{-1}$ and gives an estimated decay rate of $9.0\times10^6$ s$^{-1}$ for the $28g$ state. Predissociation rates appear as violet bars stacked on the green autoionization rate bars.

The systematic difference of around 30\% between the calculated autoionization rates and the faster observed total decay rates appears clearly in the averaged data of Panel (c) of Figure \ref{fig:ratedata_summaryfigure}. This figure also suggests a clear origin for the missing decay rate: predissociation. The absence of a state-specific model for predissociation in NO on par with our model for autoionization limits our discussion to the average rates for $ng$ states presented here. New high-resolution spectroscopy of the high-$\ell$ Rydberg states of both $v=0$ and $v=1$ manifolds may shed additional light on the relative strength and state specificity of these two decay mechanisms.

\begin{figure*}
\centering
\includegraphics{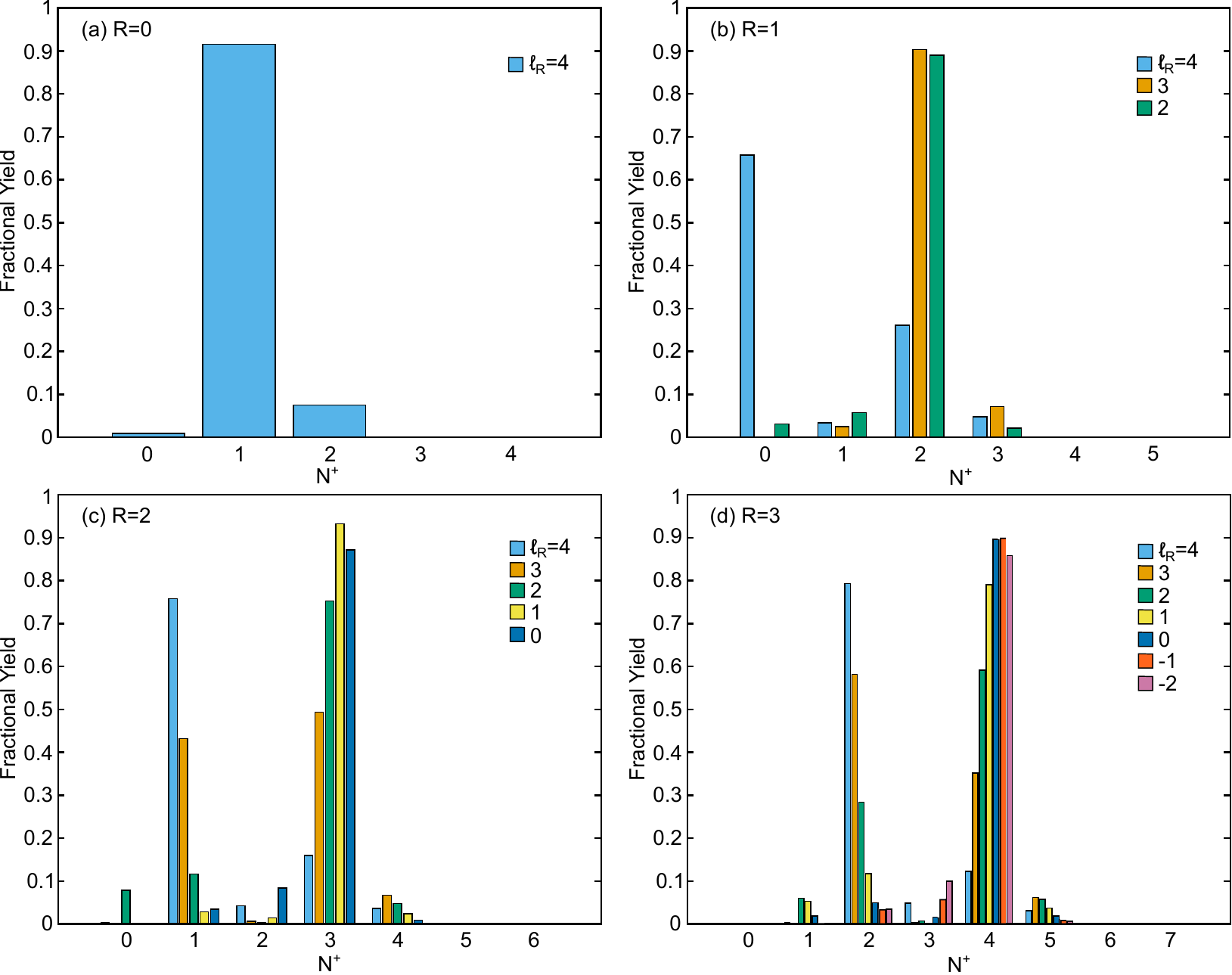}
\caption{Ion rotational state distributions following vibrational autoionization of all $\ell_R$ components of the $25g$ Rydberg state for (a) $R=0$, (b) $R=1$, (c) $R=2$, and (d) $R=3$. For $g$ states, the dipole mechanism makes the largest contribution and favors decay via $N^+-R=\pm1$ channels. Ion rotational state distributions following autoionization of a particular $\ell_R$ state are shown as colored bars according to the legend in each plot. For example, in Plot (b), the $25g1_5$ ($\ell_R=4$) state autoionizes to produce 67\% of ions in $N^+=0$, 3\% in $N^+=1$, 27\% in $N^+=2$, and 3\% in $N^+=3$, shown as light blue bars on the left-hand side for each $N^+$ value. For specific values of $\ell_R$, a greater than 90\% yield of the NO$^+$ ion in a single rotational level is observed.}
\label{fig:25giondist}
\end{figure*}

\subsection{Rotational state distributions from autoionization of $g$ Rydberg states}

Although no experimental data is available on the NO$^+$ ion rotational states accessed by vibrational autoionization of $g$ states, we have examined the predictions of our long-range model. Given the consistency between the measured total decay rates and the calculated autoionization rates, we anticipate that vibrational autoionization is the dominant decay mechanism for $g$ states and that our model accurately captures the details of the autoionization dynamics. Figure \ref{fig:25giondist} shows the calculated ion rotational state distributions for all $\ell_R$ components of the $25g$ Rydberg states with (a) $R=0$, (b) $R=1$, (c) $R=2$, and (d) $R=3$. The calculated rotational distributions differ significantly from those observed for $f$ states. Decay into the odd $N^+-R = \pm 1$ channels dominates over even $N^+-R$ channels. Of particular interest, several of the individual $\ell_R$ states (e.g., $\ell_R=3,1,-1$ for $R=1,2,3$) display a greater than 90\% yield of the NO$^+$ ion in a single rotational state. 

\begin{figure}
\centering
\includegraphics{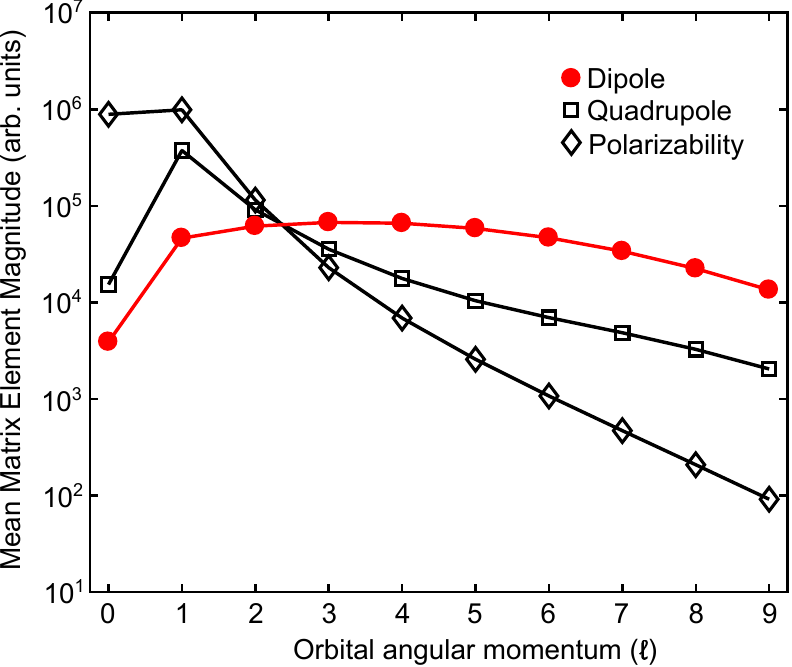}
\caption{Matrix elements calculated according to Eq.\ \ref{dipolematrixelement} for the dipole, quadrupole, and polarizability mechanisms averaged over all $\ell_R$ components and all $\ell$, $N^+$ decay channels of the $n=25$ Rydberg state as a function of the Rydberg electron orbital angular momentum $\ell$. For $d$ and $f$ states, all three mechanisms contribute similarly to vibrational autoionization resulting in population of many ion rotational states. As the value of $\ell$ increases the magnitudes of the quadrupole and polarizability mechanisms decrease rapidly. The long-range dipole mechanism dominates the decay dynamics for all high-$\ell$ states. }
\label{fig:Lvariation}
\end{figure}

The profound difference in vibrational autoionization dynamics between $f$ and $g$ Rydberg states of NO is explained by the relative strengths of the different long-range mechanisms. Figure \ref{fig:Lvariation} shows the relative magnitudes of the matrix elements (see Eq.\ \ref{dipolematrixelement}) for the dipole, quadrupole, and polarizability mechanisms averaged over all $\ell_R$ states and all $\ell$, $N^+$ decay channels for the $n=25$ Rydberg state with different values of $\ell$. At low-$\ell$, the polarizability and quadrupole mechanisms are significantly stronger than the dipole mechanism, reflecting the unusually small dipole moment of the NO$^+$ ion-core. As the value of $\ell$ increases, the centrifugal barrier pushes the Rydberg wavefunction to larger distances, $r$, from the ion-core. Since the matrix elements for the dipole, quadrupole and polarizability decrease as $r^{-2}$, $r^{-3}$, and $r^{-4}$, respectively, the magnitudes of these matrix elements decrease at different rates with increasing $\ell$. This results in a re-ordering of magnitudes. For the $d$ ($\ell=2$) and $f$ ($\ell=3$) states, all three mechanisms contribute to autoionization decay with similar magnitude. This means that $f$ Rydberg states of NO decay by vibrational autoionization into all possible ion rotational state channels ($-2 \leq N^+ - R \leq 2$) with significant amplitude. This is reflected in the model predictions shown in Figures \ref{fig:v2ellR-101}, \ref{fig:v1ellR123}, and \ref{fig:v2ellR3}.  For $\ell \geq 4$, the dipole, the longest range mechanism, is nearly an order of magnitude stronger than the other mechanisms. As a result, the two ion rotational state channels ($N^+ - R = \pm 1$) accessed by the dipole mechanism are the dominant decay pathways. In the high rotation limit, the $\ell_R = \pm \ell$ states decay preferentially into the channels $N^+ - R = \mp 1$, while states with $\ell_R$ values closer to zero decay into both rotational channels. As the value of $R$ increases from Panel (a) to (d), this limiting behavior is approached: the yield of the $\ell_R=4$ state into the $N^+-R=-1$ channel increases and the $\ell_R$ value with the maximum $N^+-R=+1$ yield approaches $-\ell$. At low rotation, this pattern is disrupted and this hinders the production of NO$^+$ in the ground $N^+=0$ state. A maximum yield of just 0.67 is obtained from the $\ell_R=4$, $R=1$ state. All other $N^+$ states are accessible in high yield from a particular $\ell_R$ state in the Rydberg complex. For example, one could prepare a $ng2_3$ ($v=1$, $\ell_R=1$) level by the triple resonance scheme on line 8 of Table \ref{table:excitationschemes}, and this Rydberg state spontaneously decays to generate NO$^+$ ions in the $N^+=3$ level with approximately 93\% yield. Since the dipole mechanism is the dominant mechanism for all states with $\ell \geq 4$, this pattern in the vibrational autoionization decay dynamics is expected to occur for all high-$\ell$ Rydberg states of NO and all other heteronuclear diatomic molecules.

This universal vibrational autoionization behavior makes high-$\ell$ Rydberg states attractive targets for state-selective production of molecular ions. Several experiments in precision measurement\cite{loh2011,germann2014} and cold chemistry\cite{schmid2019,tong2012} require molecular ions in a single quantum state, but typical methods of ion production (e.g., discharge, laser ablation) are violent and produce ions in numerous quantum states. In contrast, autoionization of Rydberg states has been used in a few previous experiments\cite{loh2011,zhou2019} to generate molecular ions in only a few quantum states, greatly increasing the yield of the desired state and reducing the complexity of subsequent state preparation steps. The calculated rotational state distributions presented here demonstrate that greater selectivity in the generation of single quantum state ions is possible by careful selection of high-$\ell$ Rydberg states as the precursor to autoionization. New methods for the preparation of high-$\ell$ Rydberg states\cite{morgan2018,barnum2020} will be important for exploiting this approach.

\section{Conclusion}

We have examined the predictions of a long-range model for vibrational autoionization of Rydberg states of NO and have validated the simplified physical mechanisms of this model by comparison with experimental data for Rydberg states with $\ell=3$ and 4. We find that the long-range model predicts NO$^+$ rotational state distributions that result from autoionization of $f$ Rydberg states that are largely consistent with the experimental observations.\cite{park1996,park1997,zhao2004_thesis} The agreement is particularly striking for states with $|\ell_{R}|\leq1$. The extreme $\ell_{R}$ components decay by additional $N^+-R=\pm3$, and $\pm4$ channels, which are not captured by our model. Future extensions of the long-range model to include higher-order polarizabilities and explicit phase shifts due to core-penetration, or application of multichannel quantum defect theory will account for these discrepancies. Our results strongly support a direct, vibrational mechanism for autoionization, and suggest that autoionization occurs independently of the much faster predissociation in the $f$ Rydberg states.

Non-radiative lifetimes of $g$ Rydberg states of NO have been directly measured by delayed pulsed field extraction of NO$^+$ ions that result from autoionization. We find good agreement between the experimental total decay rates and the autoionization rates predicted by our long-range model, though the calculated rates are approximately 30\% slower than the observed rates. This difference between experiment and the autoionization calculation can be accounted for by predissociation of the $g$ states. In contrast to the behavior of all low-$\ell$ Rydberg states of NO, autoionization is the fastest non-radiative decay mechanism of $g$ Rydberg states.

Finally, we propose that vibrational autoionization of selected $g$ Rydberg states is an efficient strategy for producing molecular ions in single selected quantum states. While our investigation focuses on NO, this experimental and theoretical methodology will be applicable to a variety of molecules because high-$\ell$ Rydberg states obey a universal scaling of the long-range autoionization mechanisms. Efficient production of quantum-state selected ions is desirable for diverse applications in precision measurement,\cite{loh2011,germann2014} cold chemistry,\cite{schmid2019,tong2012} and quantum computing.\cite{schuster2011}

\section*{Data Availability}
The data that support the findings of this study are available from the corresponding author upon reasonable request.

% If in two-column mode, this environment will change to single-column format so that long equations can be displayed. 
%\begin{widetext}
%$$\mbox{put long equation here}$$
%\end{widetext}
%
% Figures should be put into the text as floats. 
% Use the graphics or graphicx packages (distributed with LaTeX2e).
% Put the label that you will use with \ref{} command in the braces of the \label{} command.
%
% \begin{figure}
% \end{figure}
% \begin{table}
% \end{tabular}
% \end{table}

\begin{acknowledgments}

We gratefully acknowledge insightful conversations with Ed Eyler (University of Connecticut) and Heather Lewandowski (University of Colorado/JILA) that inspired this work. This material is based on work supported by the National Science Foundation, under Award No.\ CHE-1800410 and the AFOSR, under Award No.\ FA9550-16-1-0117. T.J.B.\ was supported by the National Science Foundation Graduate Research Fellowship Program under Grant No.\ 1122374. J.J.\ assisted in the preparation of this manuscript at Lawrence Livermore National Laboratory under the auspices of the U.S. Department of Energy under Contract DE-AC52-07NA27344.
\end{acknowledgments}

\bibliography{main}

\end{document}